\documentclass{article}

\usepackage{arxiv}

\usepackage[utf8]{inputenc} 
\usepackage[T1]{fontenc}    
\usepackage{hyperref}       
\usepackage{url}            
\usepackage{booktabs}       
\usepackage{amsfonts}       
\usepackage{nicefrac}       
\usepackage{microtype}      
\usepackage{lipsum}

\usepackage{cite}
\usepackage{amsmath,amssymb,amsfonts,amsbsy}
\usepackage{algorithmic}
\usepackage{graphicx,xcolor}
\usepackage{textcomp}
\usepackage{xcolor}
\usepackage{subfigure} 
\usepackage{adjustbox}
\usepackage{comment}

\usepackage{epstopdf} 
\usepackage{adjustbox} 

\usepackage[]{algorithm2e}
\newcommand{\hl}[1]{\textcolor{red}{#1}}

\DeclareMathOperator*{\argmax}{arg\,max}

\title{Epileptic seizure prediction using Pearson's product-moment correlation coefficient of a linear classifier from generalized Gaussian modeling}

\author{
  Antonio Quintero-Rinc\'on, Carlos D'Giano\\
  Centro Integral de Epilepsia y Telemetr\'ia\\
  Fundaci\'on Lucha contra las Enfermedades Neurol\'ogicas Infantiles (FLENI)\\
  Buenos Aires, Argentina \\
  \texttt{tonioquintero@ieee.org, cdigiano@fleni.org.ar } \\
   \And
 Marcelo Risk \\
 Departmento de Bioingenier\'ia\\
 Instituto Tecnol\'ogico de Buenos Aires (ITBA)\\
 Buenos Aires, Argentina \\
 \texttt{mrisk@itba.edu.ar} }

\begin{document}
\maketitle

\begin{abstract}
Predecir una crisis epil\'eptica, significa la capacidad de determinar de antemano el momento de una crisis con la mayor precisi\'on posible. Un pron\'ostico correcto de un evento epil\'eptico en aplicaciones cl\'inicas es un problema t\'ipico en procesamiento de se\~nales biom\'edicas, lo cual ayuda a un diagn\'ostico y tratamiento apropiado de esta enfermedad. En este trabajo, utilizamos el coeficiente de correlaci\'on producto-momento de Pearson a partir de las clases de un clasificador lineal estimandos usando los par\'ametros de la distribuci\'on Gaussiana generalizada. Esto con el fin de poder pronosticar eventos con crisis y eventos con no-crisis en se\~nales epil\'epticas. El desempe\~no en 36 eventos epil\'epticos de 9 pacientes, muestran un buen rendimiento con un 100\% de efectividad para sensibilidad y especificidad superior al 83\% para eventos con crisis en todos los ritmos cerebrales. El test de Pearson sugiere que todos los ritmos cerebrales est\'an altamente correlacionados en los eventos con no-crisis, m\'as no durante los eventos con crisis. Esto sugiere que nuestro modelo puede escalarse con el coeficiente de correlaci\'on  producto-momento de Pearson para la detecci\'on de crisis en se\~nales epil\'epticas.\\

To predict an epileptic event, means the ability to determine in advance the time of the seizure with the highest possible accuracy. A correct prediction benchmark for epilepsy events in clinical applications, is a typical problem in biomedical signal processing that help to an appropriate diagnosis and treatment of this disease. In this work we use Pearson's product-moment correlation coefficient from generalized Gaussian distribution parameters coupled with linear-based classifier to predict between seizure and  non-seizure events in epileptic EEG signals. The performance in 36 epileptic events from 9 patients showing a good performance with 100\% of effectiveness for sensitivity and  specificity greater than 83\% for seizures events in all brain rhythms. Pearson's test suggest that all brain rhythms are highly correlated in non-seizure events but no during the seizure events. This suggests that our model can be scaled with the Pearson product-moment correlation coefficient for the detection of epileptic seizures.
\end{abstract}


\section{Introducci\'on}

El diagn\'ostico y tratamiento adecuado de la epilepsia es uno de los  principales problemas de la salud p\'ublica seg\'un la Organizaci\'on Mundial de la Salud. En todo el mundo hay m\'as de 50 millones de personas que padecen alg\'un tipo de epilepsia, casi el 80 \% de ellas en regiones en desarrollo, donde tres cuartas partes no reciben un diagn\'ostico y tratamiento apropiado \cite{WHO2018}. Los pacientes que padecen esta enfermedad a menudo manifiestan diferentes caracterizaciones fisiol\'ogicas, que resultan de la descarga sincr\'onica y excesiva de un grupo de neuronas en la corteza cerebral. Las crisis epil\'epticas generalmente tienen un inicio repentino, se extienden en cuesti\'on de segundos y, en la mayor\'ia de los casos son breves. La manifestaci\'on de una crisis depende de la regi\'on d\'onde comienza en el cerebro y qu\'e tan r\'apido se propaga. La correcta identificaci\'on de esta informaci\'on es clave para un tratamiento adecuado de esta enfermedad.

La electroencefalograf\'ia (EEG) es una modalidad biom\'edica no invasiva y ampliamente disponible que se puede utilizar para diagnosticar y dise\~nar un tratamiento correcto de la epilepsia. El EEG captura las principales caracter\'isticas que son relevantes de la crisis, lo cual ayuda a discriminar entre actividad cerebral normal y anormal. Las caracter\'isticas m\'as estudiadas en la literatura se pueden clasificar en tres grupos: propiedades espectrales, propiedades morfol\'ogicas y descriptores estad\'isticos. Para un tratamiento integral de estas caracter\'isticas ver \cite{NiedermeyerDaSilva2010,EpilepsyIntersection2011, EpilepticSeizures2010}.

Primero vamos a explicar brevemente algunos conceptos usados en este trabajo para desarrollar toda la idea propuesta.

La \emph{validaci\'on cruzada} es una t\'ecnica de validaci\'on de modelos usada para evaluar c\'omo los resultados de un algoritmo de an\'alisis estad\'istico se pueden generalizar a un conjunto de datos independiente. Esto se hace mediante la partici\'on de un conjunto de datos de la siguiente manera: un subconjunto para entrenar el algoritmo y los datos restantes para la prueba o test. Cada ronda de validaci\'on cruzada implica la partici\'on aleatoria del conjunto de datos original en un \emph{conjunto de entrenamiento} y un \emph{conjunto de test}. El \emph{conjunto de entrenamiento} luego se usa para entrenar un algoritmo de aprendizaje supervisado y el \emph{conjunto de test} se usa para evaluar su desempe\~no. Este proceso se repite varias veces donde el \emph{valor de p\'erdida y el error aparente} de la validaci\'on cruzada se utilizan como un indicador de rendimiento. Aplicaciones de este m\'etodo en epilepsia se remontan a la d\'ecada de los 70's con \cite{Lloyd1972} usando una t\'ecnica llamada \emph{template matching} \cite{QuinteroRincon2016a}. La importancia de usar la validaci\'on cruzada radica en que en muchas aplicaciones biom\'edicas los datos pueden ser muy l\'imitados  para la etapa de entrenamiento y de test, por lo que si se quieren construir buenos modelos, se debe utilizar la mayor cantidad de datos disponibles para la etapa de entrenamiento. Sin embargo, si el conjunto de validaci\'on es peque\~no, dar\'a una estimaci\'on relativamente ruidosa del rendimiento predictivo \cite{Bishop2006}. En la \emph{Validaci\'on cruzada dejando uno fuera}, los datos de las particiones usan el enfoque de $k$-iteraciones,  donde $k$ es igual al n\'umero total de observaciones en los datos. Ver \cite{Alpaydin2014,Hastie2011} para un tratamiento exhaustivo de las propiedades estad\'isticas y \cite{Combrisson2015,Sargolzaei2015,Zhang2014,Stevenson2014,Liang2013} para algunos ejemplos en se\~nales EEG.

El coeficiente de correlaci\'on producto-momento de Pearson es un test de asociaci\'on entre parejas de datos, es usado como una medida del grado de correlaci\'on lineal o dependencia entre dos variables, \emph{crisis} y \emph{no-crisis} en nuestro caso. El coeficiente se calcula como el cociente entre la covarianza de las dos variables y el producto de sus desviaciones t\'ipicas. Referimos al lector a \cite{Glantz2011} para un tratamiento comprensivo de este coeficiente.


En este trabajo estudiamos el coeficiente de correlaci\'on producto-momento de Pearson para predecir entre los eventos de crisis y no-crisis, a partir de una clasificaci\'on lineal de la estimaci\'on de los par\'ametros de la distribuci\'on Gaussiana generalizada, modelo estudiado en nuestros trabajos previos \cite{QuinteroRincon2014, QuinteroRincon2016a, QuinteroRincon2016b, QuinteroRincon2017,QuinteroRincon2018a}. Esta distribuci\'on tiene dos par\'ametros: \emph{escala} $\mathcal{A}$ y \emph{forma} $\mathcal{B}$ que se estiman en cada ritmo cerebral, a partir de una descomposic\'on wavelet. Por lo tanto, tenemos un conjunto de par\'ametros $\mathcal{A}$ y $\mathcal{B}$ tanto para eventos de crisis como para los eventos de no-crisis. Estos par\'ametros se clasifican a trav\'es de un clasificador lineal en dos clases:  \emph{crisis} o \emph{no-crisis}. A continuaci\'on el aporte de este trabajo, se estima un coeficiente de correlaci\'on producto-momento de Pearson para cada clase; permitiendo un rango de magnitud entre $[-1,+1]$. Este escalamiento facilita una predicci\'on de la crisis epil\'eptica en se\~nales EEG.

Este documento est\'a estructurado de la siguiente manera. La secci\'on \ref{sec:meth} describe la metodolog\'ia propuesta que se usa para describir se\~nales de EEG y discriminar entre un evento de crisis y no-crisis en se\~nales epil\'epticas de EEG. Esta metodolog\'ia se aplica y luego se compara con dos modelos similares en se\~nales reales de EEG de pacientes que sufren crisis epil\'epticas en la secci\'on \ref{sec:results}. La elecci\'on de estos dos modelos es porque usan una metodolog\'ia similar y est\'an basados en el cl\'asico clasificador de m\'aquinas de vectores de soporte (SVM). Finalmente las conclusiones se informan en la secci\'on \ref{sec:disc}.

\section{Metodolog\'ia}
\label{sec:meth}
Sea $\boldsymbol X \in \mathbb{R}^{N\times M}$ la matriz en conjunto de $M$ se\~nales EEG $\boldsymbol{x}_m \in \mathbb{R}^{N\times 1}$, medidas simultaneamente en diferentes canales en instantes de tiempo discretos $N$. La metodolog\'ia propuesta esta compuesta de 5 estapas. 

La primera etapa divide la se\~nal original $\boldsymbol X$ en una serie de segmentos de 2 segundos con $50\%$ de solapamiento, usando una ventana rect\'angular $\boldsymbol \Omega = \boldsymbol\Omega_0 \left( w-\frac{W-1}{2}\right)$ con $0 \leq w\leq W-1$, tal que $\boldsymbol X^{(i)} = \boldsymbol \Omega^{(i)} \boldsymbol X$. 
La segunda etapa consiste en representar cada segmento $\boldsymbol X^{(i)}$  en su correspondiente representaci\'on tiempo-frecuencia usando una descomposici\'on multiresoluci\'on 1D, a trav\'es de la wavelet Daubechies (dB4) con 6 escalas. El prop\'osito de esta descomposici\'on es evaluar la distribuci\'on de energ\'ia  a trav\'es de todos los ritmos cerebrales llamados: \emph{banda delta: 0.5-4Hz, banda theta:4-8Hz, banda alfa: 8-13Hz, banda beta:13-30Hz} y \emph{banda gamma: $>$ 30Hz}. 
\begin{align}
	\boldsymbol X^{(i)}_b &= \left [\boldsymbol X^{(i)}_{\delta} \; \boldsymbol X^{(i)}_{\theta}\; \boldsymbol X^{(i)}_{\alpha} \; \boldsymbol X^{(i)}_{\beta} \;  \boldsymbol X^{(i)}_{\gamma} \right ]^T
\end{align}

En la tercera etapa, la distribuci\'on estad\'istica de los coeficientes wavelet es representada usando la distribuci\'on generalizada Gaussiana (GGD) de media cero, estudiada en nuestros trabajos previos \cite{QuinteroRincon2014,QuinteroRincon2016a, QuinteroRincon2016b, QuinteroRincon2017,QuinteroRincon2018a}.  La GGD tiene una funci\'on de densidad de probabilidad (PDF) dada por:
\begin{equation}
	\label{eq:ggd}
	f_\textnormal{GGD}(x;\mathcal{A},\mathcal{B}) = \frac{\mathcal{B}}{2\mathcal{A}\Gamma(\mathcal{B}^{-1})} \exp\left(-\left|\frac{x}{\mathcal{A}}\right|^\mathcal{B}\right)
\end{equation}
donde $\mathcal{A} \in \mathbb{R}^+$ es el par\'ametro de \emph{escala}, $\mathcal{B} \in \mathbb{R}^+$ es el par\'ametro de \emph{forma} y $\Gamma\left(\cdot\right)$ es la funci\'on Gamma. 

Cada escala de la descomposici\'on wavelet es reducida al estimar los par\'ametros estad\'isticos de la distribuci\'on Gaussiana generalizada $\mathcal{A}$ y $\mathcal{B}$, (Ver ecuaci\'on \eqref{eq:ggd}), con el fin de obtener el conjunto de caracter\'isticas asociadas a todas las escalas wavelet, para un segmento de 2 segundos con $50\%$ de solapamiento.

\begin{align}
	\boldsymbol {\widehat X}^{(i)}_{b} &= \left [\boldsymbol X^{(i)}_{(\mathcal{A},\mathcal{B}),{\delta}} \; \boldsymbol X^{(i)}_{(\mathcal{A},\mathcal{B}),{\theta}} \; \boldsymbol X^{(i)}_{(\mathcal{A},\mathcal{B}),{\alpha}} \; \boldsymbol X^{(i)}_{(\mathcal{A},\mathcal{B}),{\beta}} \; \boldsymbol X^{(i)}_{(\mathcal{A},\mathcal{B}),{\gamma}} \right]^T 
	= \argmax_{\left[\mathcal{A},\mathcal{B}\right]^T} f_\textnormal{GGD}\left (\boldsymbol X^{(i)}_b; \mathcal{A},\mathcal{B} \right ) 
\end{align}
En la cuarta etapa se utiliza un an\'alisis discriminante lineal para clasificar en dos clases posibles: $\omega_s$ para los eventos de crisis y $\omega_{ns}$ para los eventos de no-crisis. Para un vector de caracter\'isticas $\boldsymbol {\widehat X}^{(i)}_b$ perteneciente a la clase $\omega_s$ o a la clase $\omega_{ns}$, se asume que $\boldsymbol {\widehat X}^{(i)}_b$ tiene una distribuci\'on normal con valor medio $\boldsymbol\mu_s$ (o $\boldsymbol\mu_{ns}$) y matriz de covarianza $\boldsymbol\Sigma_s$ (o $\boldsymbol\Sigma_{ns}$), entonces:
\begin{align}
	\label{eq:linear1}
	P \left(\boldsymbol {\widehat X}^{(i)}_b \middle|\omega_{s}\right) &= \frac{1}{\sqrt{ \left( 2 \pi \right)^k \left| \boldsymbol\Sigma_s \right| }} \exp \left[ -\frac12 \left( \boldsymbol {\widehat X}^{(i)}_b -\boldsymbol\mu_s \right)^T \boldsymbol\Sigma_s^{-1} \left( \boldsymbol {\widehat X}^{(i)}_b - \boldsymbol\mu_s \right)  \right] \\
	P \left(\boldsymbol {\widehat X}^{(i)}_b \middle|\omega_{ns}\right) &= \frac{1}{\sqrt{ \left( 2 \pi \right)^k \left| \boldsymbol\Sigma_{ns} \right| }} \exp \left[ -\frac12 \left( \boldsymbol {\widehat X}^{(i)}_b - \boldsymbol\mu_{ns} \right)^T \boldsymbol\Sigma_{ns}^{-1} \left( \boldsymbol {\widehat X}^{(i)}_b - \boldsymbol\mu_{ns} \right)  \right]
	\label{eq:linear2}	
\end{align}
donde $k$ es la dimensi\'on del vector estimado $\boldsymbol {\widehat X}^{(i)}_b$ y $P(\cdot)$ es la probabilidad de evento en particular. 
Para el an\'alisis discriminante lineal, se calculan las muestras $\boldsymbol\mu_s$ (o $\boldsymbol\mu_{ns}$) de cada clase. Entonces se calcula la muestra $\boldsymbol\Sigma_s$ (o $\boldsymbol\Sigma_{ns}$)  al restar primero la muestra $\boldsymbol\mu_s$ (o $\boldsymbol\mu_{ns}$) de cada clase a partir de las observaciones de esa clase, y tomando la matriz emp\'irica $\boldsymbol\Sigma_s$ (o $\boldsymbol\Sigma_{ns}$) del resultado. Por lo tanto el discriminante lineal para el problema de clasificaci\'on viene dado por 
\begin{align}
	\log \frac{ \rho \left(\boldsymbol {\widehat X}^{(i)}_b \middle|\omega_{s}\right) }{ \rho \left(\boldsymbol {\widehat X}^{(i)}_b \middle|\omega_{ns}\right) }
	&= (\boldsymbol {\widehat X}^{(i)}_b - \boldsymbol\mu_s)^T \boldsymbol\Sigma_s^{-1}(\boldsymbol {\widehat X}^{(i)}_b - \boldsymbol\mu_s) + ln \big | \boldsymbol\Sigma_s \big |  \nonumber \\
	&~~~~ - (\boldsymbol {\widehat X}^{(i)}_b - \boldsymbol\mu_{ns})^T \boldsymbol\Sigma_{ns}^{-1}(\boldsymbol {\widehat X}^{(i)}_b - \boldsymbol\mu_{ns}) - ln \big | \boldsymbol\Sigma_{ns} \big |
	\label{eq:Bayes}.	
\end{align}

Finalmente, en la etapa cinco, el coeficiente $r$ de correlaci\'on producto-momento de Pearson  se estima a trav\'es de
\begin{align}
	r = \frac{\sum(\omega_{s} - \overline{\omega_{s}})(\omega_{ns} - \overline{\omega_{ns}})} {\sqrt{\sum(\omega_{ns} - \overline{\omega_{s}})^2\sum(\omega_{ns}- \overline{\omega_{ns}})^2}}
	\label{eq:pearson}	
\end{align}
donde $\overline{\omega_{s}}$ y $\overline{\omega_{ns}}$ son las medias de cada clase. La magnitud de $r$ describe la fuerza de asociaci\'on entre las dos variables y el signo de $r$ indica la direcci\'on de esta asociaci\'on: $r=+1 $ cuando las dos variables aumentan juntas, y $r=-1$ cuando una disminuye y la otra aumenta. As\'i mismo, tambi\'en muestra el caso m\'as com\'un de dos variables que est\'an correlacionadas linealmente. El valor $r=0$ indica ausencia de correlaci\'on, $r=+1$ indica correlaci\'on positiva total  y $r=-1$ indica correlaci\'on negativa total.

\section{Resultados}
\label{sec:results}

La metodolog\'ia propuesta se evalu\'o mediante la base de datos del Hospital Infantil de Boston, que consta de 36 registros de EEG de sujetos pedi\'atricos con crisis intratables. Las se\~nales EEG son bipolares y est\'an muestreadas a 256Hz para cada sujeto. Cada registro contiene un evento de crisis con un inicio y un final marcado, el cual fue detectado por un neur\'ologo experimentado. En este trabajo usamos 18 eventos de crisis y 18 eventos de no-crisis de 9 sujetos. Consulte \cite{Goldberger2000} para obtener m\'as detalles.

Durante la etapa de pre-propresamiento se usaron dos filtros Butterworth IIR en cascada, un filtro pasa-bajo de segundo orden con frecuencia de corte de 100 Hz y un filtro pasa-alto de primer orden con una frecuencia de corte de 30 Hz, adem\'as se sustrajo el valor medio de cada canal. Consultar \cite{QuinteroRincon2012} para un amplio estado del arte en diferentes tipos de artefactos en se\~nales EEG.

La detecci\'on de una crisis consta de dos etapas principales: la extracci\'on de caracter\'isticas y una etapa de clasificaci\'on basado en un aprendizaje autom\'atico, con el fin de caracterizar y cuantificar eventos de crisis o eventos de no-crisis. Usando los mismos datos de entrada, nuestro modelo \textbf{[Q]} fue comparado con dos modelos similares del estado del arte que trabajan tambi\'en sobre todos los ritmos cerebrales con una longitud de ventana de 2 segundos y solapamiento del 50\%: \textbf{[S]} Shoeb et al \cite{Shoeb2004,Shoeb2010} Usando una ventana rect\'angular, la extracci\'on de caracter\'isticas se realiza a trav\'es del calculo de las diferencias de energ\'ia a nivel espectral/espacial y su relaci\'on espectral/temporal
utilizando una wavelet Daubechies (dB4) con 6 escalas junto con la densidad espectral de potencia. \textbf{[C]} Chan et al. \cite{Chan2008} Usando una ventana Hamming, la extracci\'on de caracter\'isticas se estiman a trav\'es del espectro de potencia usando la transformada de Fourier (FFT) junto con un  periodograma. 
Ambos modelos usan un clasificador basado en m\'aquinas de vectores de soporte (SVM).  Cabe resaltar que en esta comparaci\'on a pesar de que la extracci\'on de caracter\'isticas tiene algunas diferencias y la etapa de clasificaci\'on es distinta,  permite contrastar metodolog\'ias similares y el costo computacional que puede ser crucial en implementaciones en tiempo real. Por ejemplo una soluci\'on \'optima para el cl\'asico clasificador SVM implica una complejidad del orden de $n^2$ o $n^3$ productos, donde $n$ es el tama\~no del  conjunto de datos, el cual por lo general es grande cuando se analizan se\~nales EEG \cite{Bordes2005,ShalevShwartz2008}, mientras que para un clasificador lineal la complejidad es del orden $mn + mt + nt$, donde $m$ es el n\'umero de muestras, $n$ es el n\'umero de caracter\'isticas y $t = min(m, n)$ \cite{Cai2008}.

\begin{figure}[htbp]
	\centering
	\subfigure[\text{[Q]}:Clasificador lineal]{\includegraphics[width=52mm]{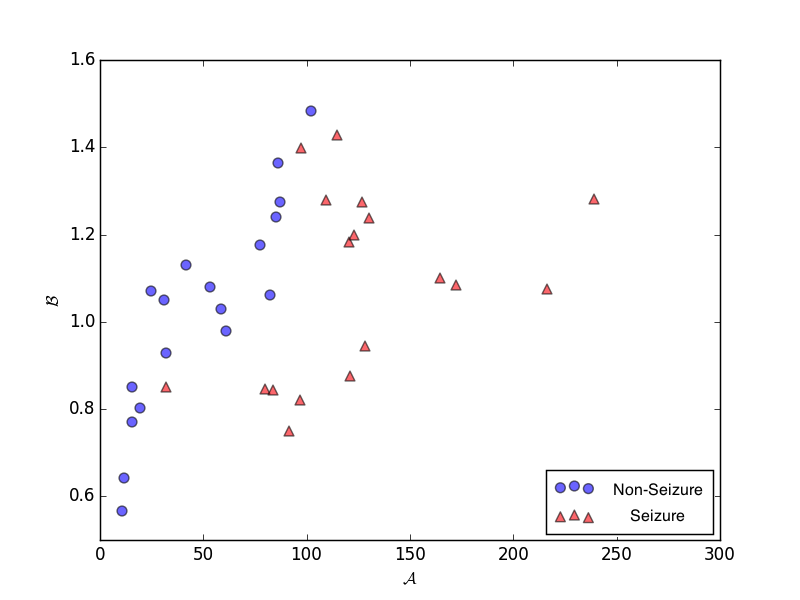}}
	\subfigure[\text{[S]}:Clasificador SVM]{\includegraphics[width=52mm]{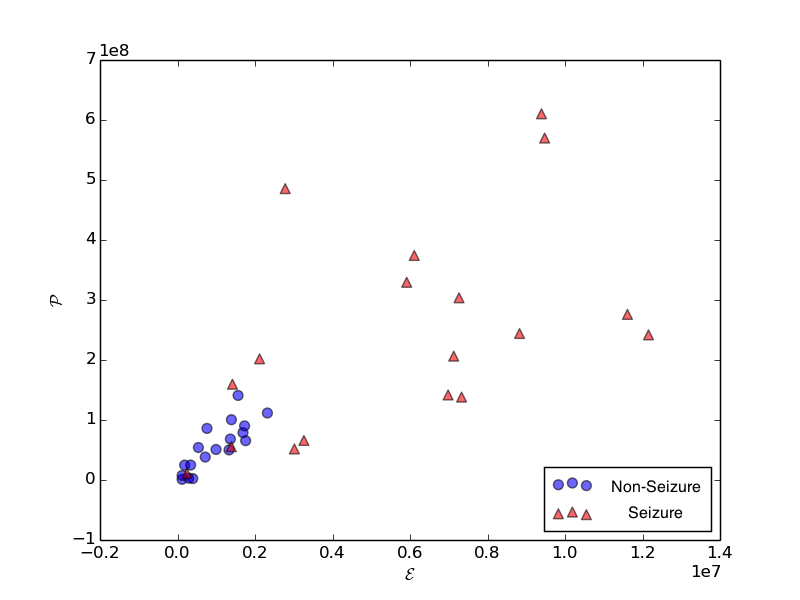}}
	\subfigure[\text{[C]}:Clasificador SVM]{\includegraphics[width=52mm]{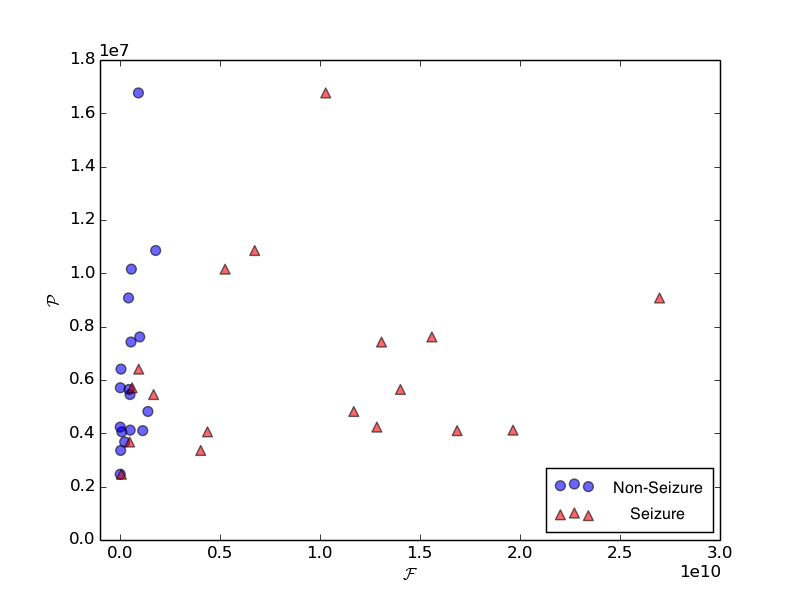}}
	\caption{Diagrama de dispersi\'on de la banda delta: (a) [Q] escala $\mathcal{A}$ eje-$x$ y forma $\mathcal{B}$ eje-$y$. (b) [S] Energ\'ia $\mathcal{E}$ eje-$x$ y Potencia $\mathcal{P}$ eje-$y$. (c) [C] Frecuencia $\mathcal{F}$ eje-$x$ y Potencia $\mathcal{P}$ eje-$y$. Todos los m\'etodos permiten una discriminaci\'on entre eventos  de no-crisis (circulos azules (\emph{non-seizure})) y eventos de crisis (tri\'angulos rojos (\emph{seizure})).} 
	\label{fig:Deltas}
\end{figure}

\begin{figure}[htbp]
	\centering
	\subfigure[\text{[Q]}:Clasificador lineal]{\includegraphics[width=52mm]{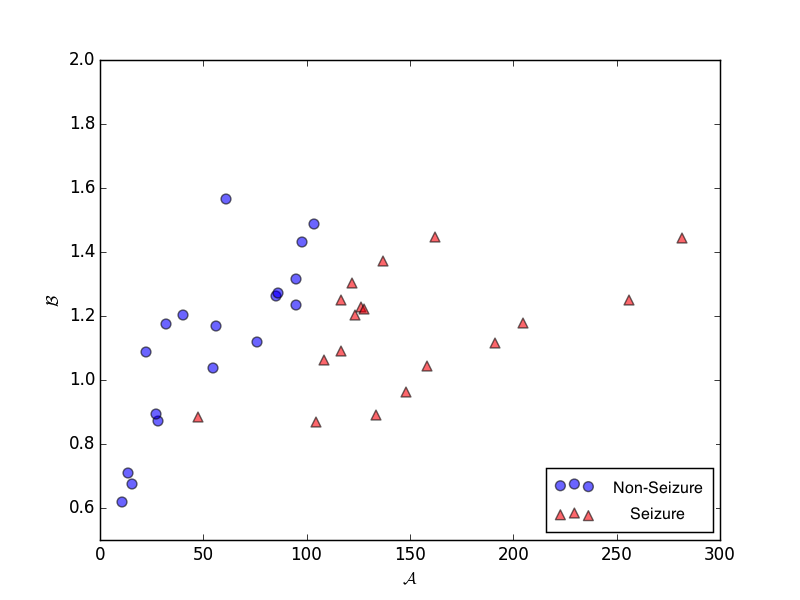}}
	\subfigure[\text{[S]}:Clasificador SVM]{\includegraphics[width=52mm]{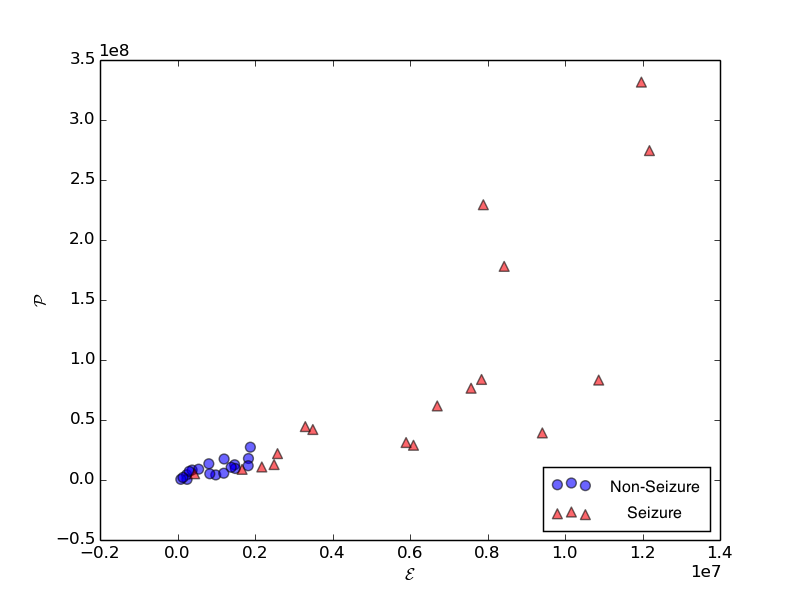}}
	\subfigure[\text{[C]}:Clasificador SVM]{\includegraphics[width=52mm]{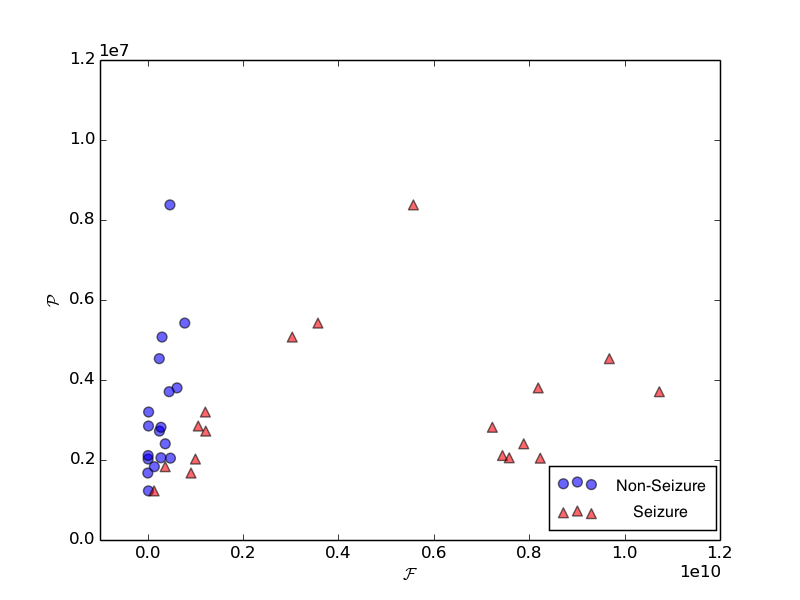}}
	\caption{Diagrama de dispersi\'on de la banda theta: (a) [Q] escala $\mathcal{A}$ eje-$x$ y forma $\mathcal{B}$ eje-$y$. (b) [S] Energ\'ia $\mathcal{E}$ eje-$x$ y Potencia $\mathcal{P}$ eje-$y$. (c) [C] Frecuencia $\mathcal{F}$ eje-$x$ y Potencia $\mathcal{P}$ eje-$y$. Todos los m\'etodos permiten una discriminaci\'on entre eventos  de no-crisis (circulos azules (\emph{non-seizure})) y eventos de crisis (tri\'angulos rojos (\emph{seizure})).} 
	\label{fig:Thetas}
\end{figure}

\begin{figure}[htbp]
	\centering
	\subfigure[\text{[Q]}:Clasificador lineal]{\includegraphics[width=52mm]{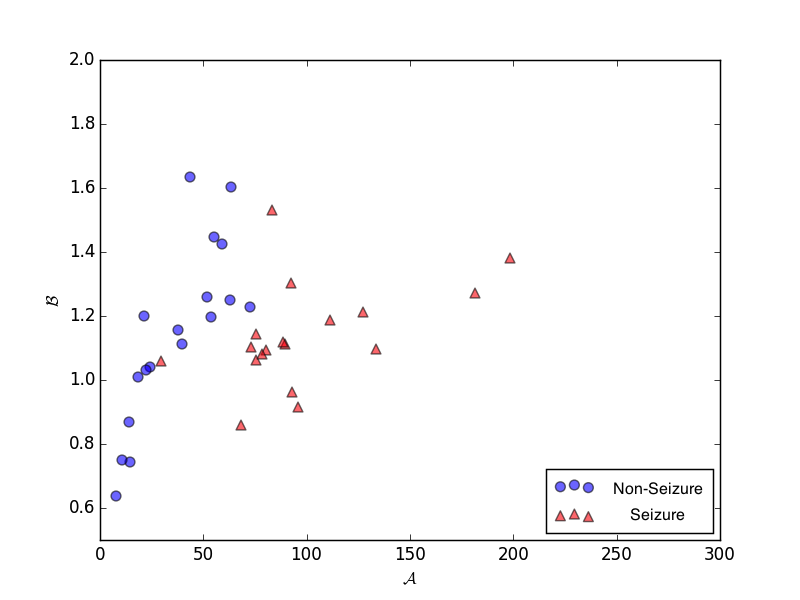}}
	\subfigure[\text{[S]}: Clasificador SVM]{\includegraphics[width=52mm]{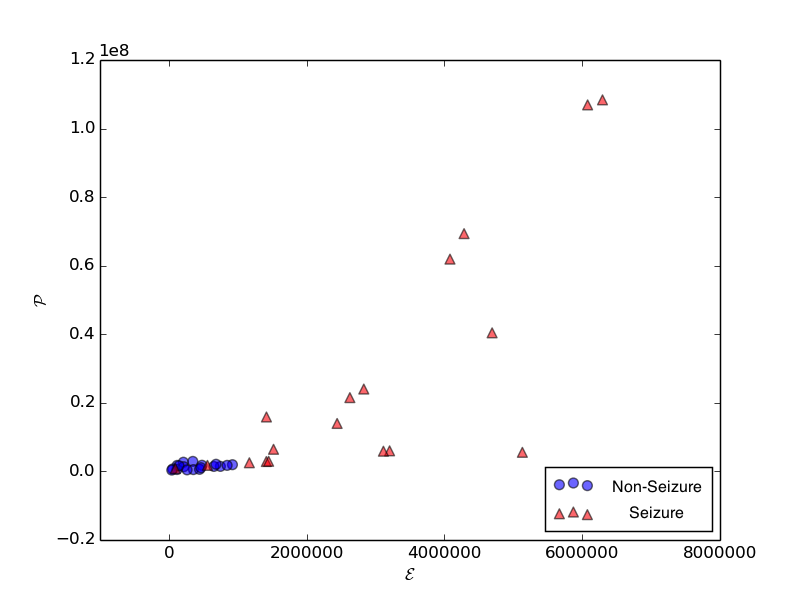}}
	\subfigure[\text{[C]}:Clasificador SVM]{\includegraphics[width=52mm]{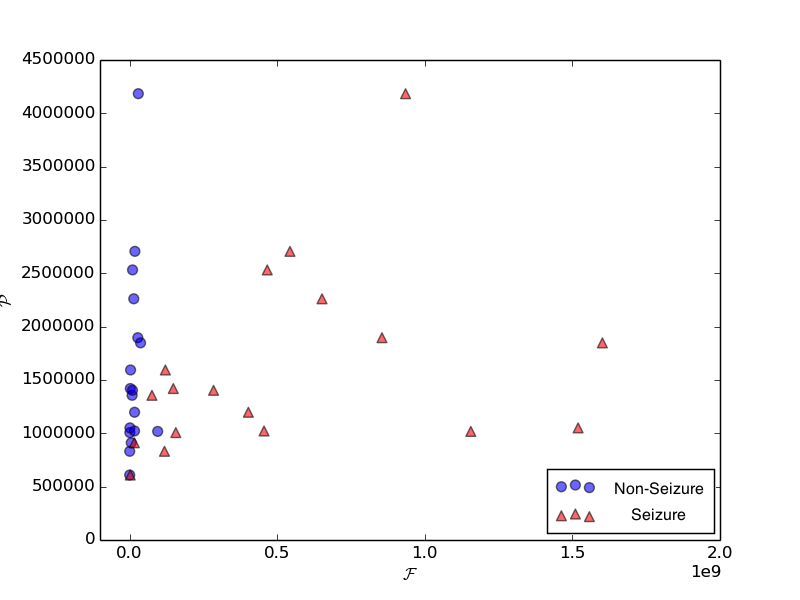}}
	\caption{Diagrama de dispersi\'on de la banda alfa: (a) [Q] escala $\mathcal{A}$ eje-$x$ y forma $\mathcal{B}$ eje-$y$. (b) [S] Energ\'ia $\mathcal{E}$ eje-$x$ y Potencia $\mathcal{P}$ eje-$y$. (c) [C] Frecuencia $\mathcal{F}$ eje-$x$ y Potencia $\mathcal{P}$ eje-$y$. Todos los m\'etodos permiten una discriminaci\'on entre eventos de no-crisis (circulos azules (\emph{non-seizure})) y eventos de crisis (tri\'angulos rojos (\emph{seizure})).} 
	\label{fig:Alphas}
\end{figure}

\begin{figure}[htbp]
	\centering
	\subfigure[\text{[Q]}:Clasificador lineal]{\includegraphics[width=52mm]{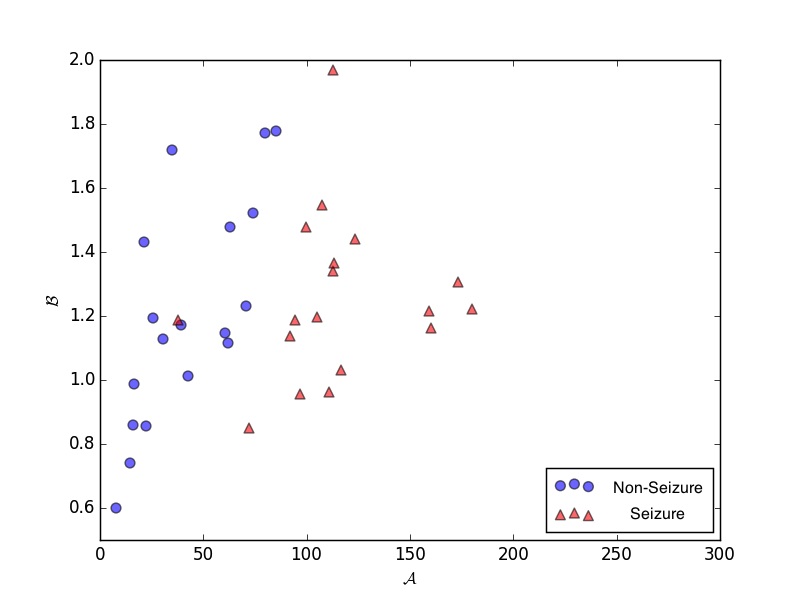}}
	\subfigure[\text{[S]}:Clasificador SVM]{\includegraphics[width=52mm]{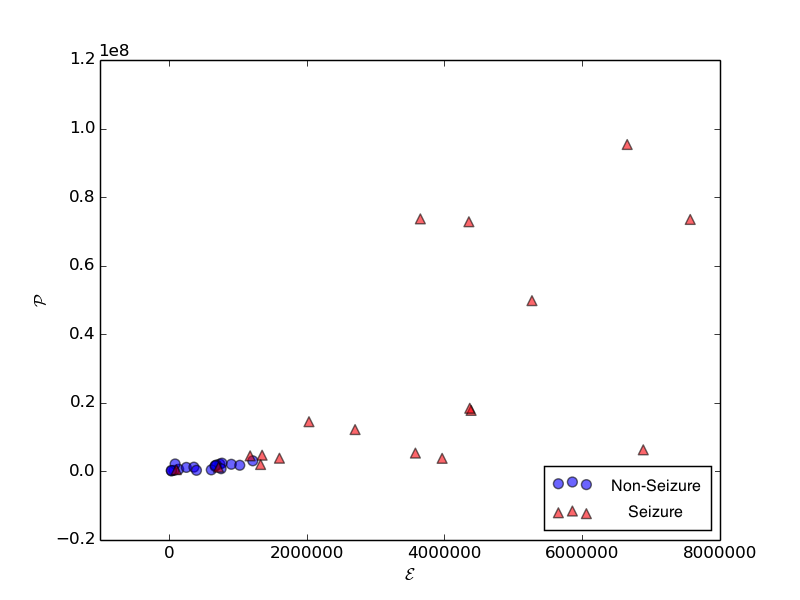}}
	\subfigure[\text{[C]}:Clasificador SVM]{\includegraphics[width=52mm]{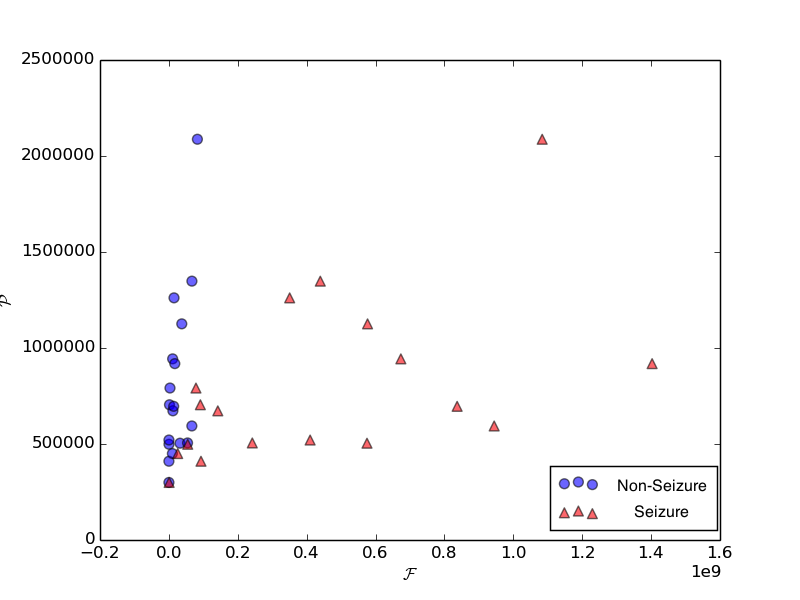}}
	\caption{Diagrama de dispersi\'on de la banda beta: (a) [Q] escala $\mathcal{A}$ eje-$x$ y forma $\mathcal{B}$ eje-$y$. (b) [S] Energ\'ia $\mathcal{E}$ eje-$x$ y Potencia $\mathcal{P}$ eje-$y$. (c) [C] Frecuencia $\mathcal{F}$ eje-$x$ y Potencia $\mathcal{P}$ eje-$y$. Todos los m\'etodos permiten una discriminaci\'on entre eventos de no-crisis (circulos azules (\emph{non-seizure})) y eventos de crisis (tri\'angulos rojos (\emph{seizure})).} 
	\label{fig:Betas}
\end{figure}

\begin{figure}[htbp]
	\centering
	\subfigure[\text{[Q]}:Clasificador lineal]{\includegraphics[width=52mm]{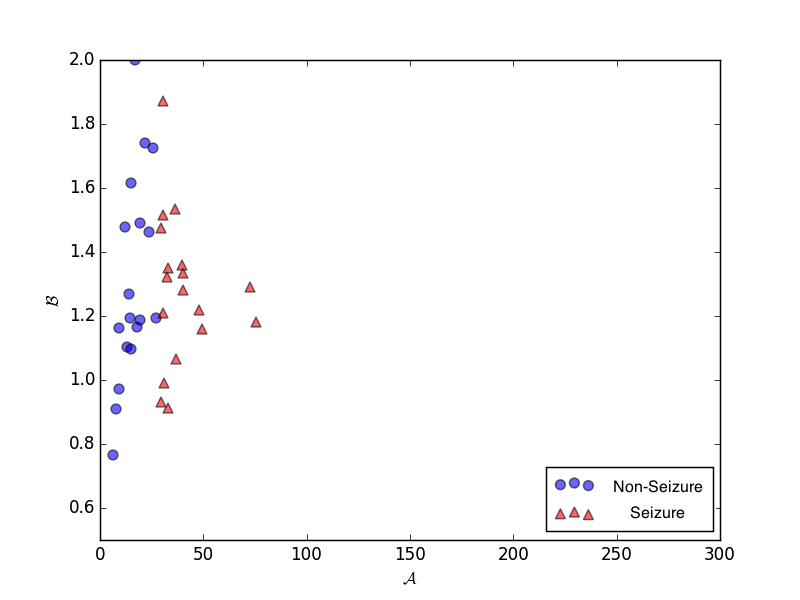}}
	\subfigure[\text{[S]}:Clasificador SVM]{\includegraphics[width=52mm]{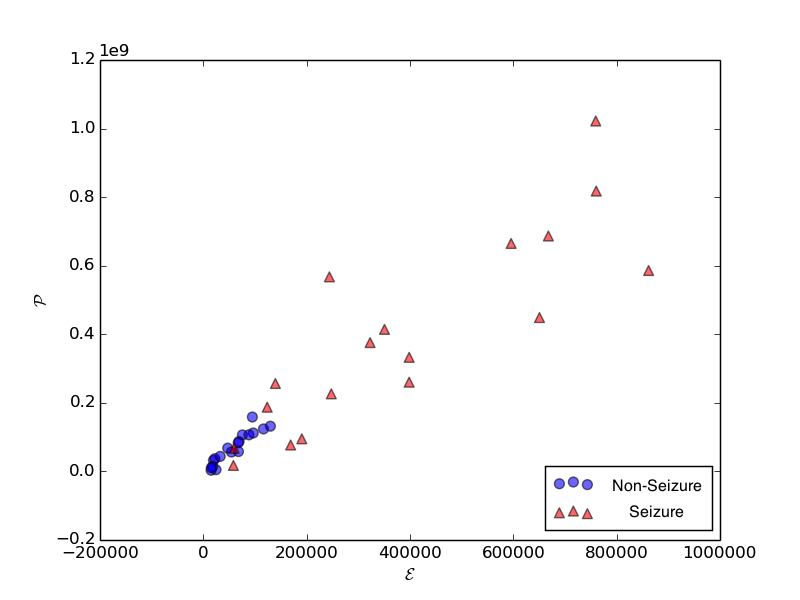}}
	\subfigure[\text{[C]}:Clasificador SVM]{\includegraphics[width=52mm]{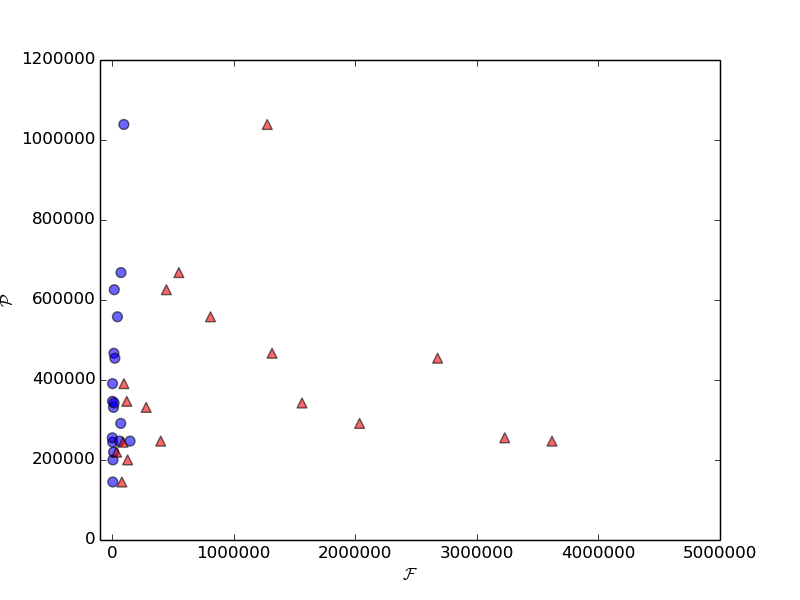}}
	\caption{Diagrama de dispersi\'on de la banda gamma: (a) [Q] escala $\mathcal{A}$ eje-$x$ y forma $\mathcal{B}$ eje-$y$. (b) [S] Energ\'ia $\mathcal{E}$ eje-$x$ y Potencia $\mathcal{P}$ eje-$y$. (c) [C] Frecuencia $\mathcal{F}$ eje-$x$ y Potencia $\mathcal{P}$ eje-$y$. Todos los m\'etodos permiten una discriminaci\'on entre eventos de no-crisis (circulos azules (\emph{non-seizure})) y eventos de crisis (tri\'angulos rojos (\emph{seizure})).} 
	\label{fig:Gammas}
\end{figure}

Las figuras \ref{fig:Deltas}-\ref{fig:Gammas} muestran el rendimiento a trav\'es de los diferentes diagramas de dispersi\'on, para todos los ritmos cerebrales de las dos clases: $\omega_s$ para crisis y $\omega_{ns}$ para no-crisis, permitiendo una buena discriminaci\'on por inspecci\'on visual para todos los modelos. La nomenclatura usada en los ejes $x$ e $y$ respectivamente, en los diagramas de dispersi\'on son: escala $\mathcal{A}$ y forma $\mathcal{B}$ para [Q] usando un clasificador lineal. Energ\'ia $\mathcal{E}$ y potencia $\mathcal{P}$ para [S] y Frequencia $\mathcal{F}$ y potencia $\mathcal{P}$ para [C], ambos usando un clasificador SVM. 

La comparaci\'on entre la tabla de contingencia o matriz de confusi\'on mostrada en la Tabla \ref{tab:confusion}, muestra una sensibilidad o porcentaje de verdaderos positivos (TPR) del 100\% para todos los modelos. Mientras que la especificidad o porcentaje de verdaderos negativos (TNR) muestra un mejor rendimiento para el modelo basado en el clasificador lineal [Q] con repecto a los otros modelos que estan basados en un clasificador SVM, [S] y [C]. Esto nos permite sugerir que nuestro modelo basado en un clasificador lineal obtiene la mejor precisi\'on para todos los ritmos cerebrales en los 36 eventos estudiados (18 no-crisis y 18 crisis). Para simplificar la interpretaci\'on visual, destacamos con color rojo el m\'etodo que logra la mayor sensibilidad, especificidad y precisi\'on general para cada banda de frecuencia.

\begin{table}[htp]
	\begin{center}
		\begin{tabular}{||c|| c|c|c|| c|c|c|| c|c|c|| c|c|c||}
			\hline \hline
			& \multicolumn{3}{|c||}{TPR} & \multicolumn{3}{|c||}{FPR} & \multicolumn{3}{|c||}{TNR} & \multicolumn{3}{|c||}{ACC} \\
			\hline 
			\hline
			Bandas		& Q & S & C & Q & S & C & Q & S & C & Q & S & C  \\
			\hline 
			\hline
			\verb Delta 	& 1 & 1& 1& \hl{0.16 }& 0.33 & 0.38 & \hl{0.83} & 0.66 & 0.61& \hl{33} & 30  & 29\\
			\verb Theta 	& 1 & 1& 1& \hl{0.05} & 0.33 & 0.38 & \hl{0.94} & 0.66 & 0.61& \hl{35} & 30  & 29 \\
			\verb Alfa 	& 1 & 1& 1 & \hl{0.11} & 0.22 & 0.38 & \hl{0.88} & 0.77 & 0.61& \hl{34} & 32  & 29\\
			\verb Beta   	& 1 & 1& 1 & \hl{0.05} & 0.33 & 0.44 & \hl{0.94} & 0.66 & 0.55 & \hl{35} & 30 & 28 \\
			\verb Gamma 	& 1 & 1& 1 & \hl{0.05} & 0.33 & 0.55 & \hl{0.94} & 0.66 & 0.44 & \hl{35} & 30 & 26 \\
			\hline \hline
		\end{tabular}
	\end{center}
	\caption{\label{tab:confusion} Comparaci\'on entre el modelo propuesto [Q] basado en un clasificador lineal, contra dos modelos similares [S] y [C] que usan un clasificador SVM. Se estudiaron 36 eventos (18 no-crisis y 18 crisis), para cada ritmo cerebral utilizando la siguiente m\'etrica: sensibilidad o porcentaje de verdaderos positivos (TPR); porcentaje de falsos positivos (FPR); especificidad o porcentaje de verdaderos negativos (TNR); y precisi\'on de clasificaci\'on general (ACC).}
\end{table}

Para cada iteraci\'on-$k$ se calculo el \emph{valor de p\'erdida} y el \emph{error aparente} de los clasificadores, para todas las observaciones  usando el modelo entrenado. El \emph{Valor de p\'erdida} permite determinar la p\'erdida de clasificaci\'on para las observaciones no utilizadas durante el entrenamiento. El \emph{error aparente} es la respuesta entre la diferencia de los datos de entrenamiento y las predicciones que hace el clasificador, en funci\'on de los datos de entrenamiento de entrada. En otras palabras es el porcentaje de error cometido sobre la misma muestra empleada para construir el modelo. Valores bajos en ambos m\'etodos significa una gran confianza o exactitud en el clasificador usado.

La Tabla \ref{tab:LRE} muestra el rendimiento de los modelos en t\'erminos de \emph{valor de p\'erdida} y el \emph{error aparente}. Se puede observar como el modelo [Q] basado en un clasificador lineal tiene los valores de p\'erdida y error m\'as bajos con respecto a los modelos [S] y [C] basados en un clasificador SVM. 

\begin{table}[htp]
	\begin{center}
		\begin{tabular}{||c|| c|c|c|| c|c|c|| c|c|c|| c|c|c||}
			\hline \hline
			& \multicolumn{3}{|c||}{Valor de p\'erdida} & \multicolumn{3}{|c||}{Error Aparente} \\
			\hline 
			\hline
			Bandas		& [Q] & [S] & [C] & [Q] & [S] & [C]  \\
			\hline 
			\hline
			\verb Delta 	& 0.083	& 0.166 & 0.250  	& 0.083 & 0.166 & 0.194 \\
			\verb Theta 	& 0.027	& 0.194 & 	0.222	& 0.027 & 0.166 & 0.194\\
			\verb Alpha 	& 0.055	& 0.138 & 	0.194	& 0.055 & 0.111 & 0.194\\
			\verb Beta   	& 0.055	& 0.166 & 	0.222	& 0.027 & 0.166 & 0.222\\
			\verb Gamma 	& 0.083	& 0.194 & 	0.277	& 0.027 & 0.166 & 0.277\\
			\hline \hline
		\end{tabular}
	\end{center}
	\caption{\label{tab:LRE} \emph{Valor de p\'erdida} y \emph{Error aparente} para los tres modelos comparados en los 36 eventos (18 no-crisis y 18 crisis), para cada ritmo cerebral. Se puede observar que los 3 modelos clasifican correctamente ya que tienen valores bajos, m\'as sin embargo nuestro modelo [Q] basado en la clasificaci\'on lineal tiene los mejores valores con respecto a [S] y [C].}
\end{table}

Teniendo toda la informaci\'on anterior, nosotros nos preguntamos si es posible  relacionar los resultados en un rango entre $[-1,+1]$ de tal forma que sirva como una de predicci\'on de las crisis. Para responder a este interrogante, se uso el coeficiente de correlaci\'on producto-momento de Pearson para las clases:  $\omega_s$ para los eventos de crisis y $\omega_{ns}$ para los eventos de no-crisis  entre el modelo [Q] basado en el clasificador lineal y los dos modelos [S] y [Q] basados en un clasificador SVM. El estudio usando 36 eventos (18 crisis y 18 no-crisis) para cada ritmo cerebral, reportado en la Tabla \ref{tab:Pearson} muestra que nuestro modelo [Q] sugiere que se pueden predecir las crisis en un rango de correlaci\'on de $[-1,+1]$ ya que, $\omega_{ns}$ para las bandas: \emph{delta},\emph{theta}, \emph{alfa} y \emph{beta} se presenta una alta correlaci\'on con un valor de $r$ cercano a $1$, excepto para la banda \emph{gamma} donde la correlaci\'on es media. Todos con un muy buen valor $p < 0.05$ permitiendo un buen intervalo de confianza. Mientras que para $\omega_{s}$ las bandas no est\'an correlacionadas ya que el valor de $r$ es cercano a cero para las bandas: \emph{delta},\emph{theta}, \emph{alfa} y \emph{beta}
mientras que para la banda \emph{gamma} la no-correlaci\'on es m\'axima. Cabe resaltar que la comparaci\'on entre las dos clases en la banda \emph{theta}, el valor de $r$ esta en la mitad del intervalo deseado para los eventos de no-crisis, m\'as no tiene un valor muy cercano respecto a los eventos de crisis. Los modelos [S] y [C] presentan en general valores de correlaci\'on muy similares, lo cual no permite discriminar claramente entre eventos de crisis y eventos de no-crisis. Estos resultados sugieren que nuestro modelo basado en la distribuci\'on Gaussiana generalizada junto con un clasificador lineal, puede definirse por un valor de escala entre $[-1,+1] $ para discriminar entre eventos de crisis y eventos de no-crisis en todas los ritmos cerebrales, permitiendo definir umbrales a partir del intervalo de confianza para la escala propuesta.

\begin{table}[htp]
	\resizebox{\textwidth}{!}{
		\centering
		\begin{tabular}{||c|| c|c|c|| c|c|c|| c|c|c|| c|c|c}
			\hline \hline
			& \multicolumn{3}{|c||}{[Q]} & \multicolumn{3}{|c||}{[C]} & \multicolumn{3}{|c||}{[S]} \\
			\hline 
			\hline
			Bandas		& r & p & IC95\% & r & p & IC95\% & r & p & IC95\% \\
			\hline 
			\hline
			Delta $\omega_{ns}$	& \hl{0.88} 	& $<$ 0.001	&  0.70  0.95	& 0.46 	& 0.054	&  0     0.76 	& 0.83	& $<$ 0.001	& 0.59 0.93 \\
			Delta $\omega_s$ & 0.39 	& 0.11		& -0.01  0.72	& 0.19 	& 0.448	& -0.30 0.60	& 0.50	& 0.034 		& 0.04 0.78  \\
			\hline
			Theta $\omega_{ns}$ & \hl{0.81} 	& $<$ 0.001	&  0.55  0.92	& 0.56 	& 0.014 	& 0.13  0.81	& 0.75	& $\approx$ 0 	& 0.45 0.90  \\
			Theta $\omega_s$ & 0.51 	& 0.03 		&  0.06  0.79	& 0.22 	& 0.363	& -0.26 0.62	& 0.76	& $\approx$ 0	& 0.46 0.90  \\
			\hline
			Alfa $\omega_{ns}$ & \hl{0.80} 	& $<$ 0.001	&  0.53  0.92	& 0.14 	& 0.558	& -0.34 0.57	& 0.33	& 0.176		& -0.15 0.69 \\
			Alfa $\omega_s$ & 0.45 	& 0.06		& -0.02  0.76	& 0.30 	& 0.217	& -0.18 0.67	& 0.80	& $<$ 0.001	& 0.53 0.92  \\
			\hline
			Beta $\omega_{ns}$ & \hl{0.72} 	& $<$ 0.001	&  0.38  0.89	& 0.60 	& 0.007	& 0.19 0.83	& 0.69     	& $<$ 0.001	& 0.34 0.87  \\
			Beta $\omega_s$ & 0.15 	& 0.56		&  -0.34 0.58	& 0.50 	& 0.031	& 0.05 0.78	& 0.65	& $\approx$ 0	& 0.26 0.85  \\
			\hline
			Gamma $\omega_{ns}$ & \hl{0.58} 	& 0.01		&  0.15  0.82	& 0.34 	& 0.167	& -.015 0.69	& 0.93     	& $\approx$ 0	& 0.82 0.97  \\
			Gamma $\omega_s$ & -0.11 	& 0.66  		&  -0.55 0.38	& 0.01	& 0.947	& -0.45 0.47	& 0.85	& $<$ 0.001	& 0.65 0.94  \\
			\hline \hline
			
	\end{tabular}}
	\caption{\label{tab:Pearson} Coeficiente de correlaci\'on producto-momento de Pearson entre $\omega_s$ para un evento de crisis (seizure) y $\omega_{ns}$ para un evento de no-crisis (non-seizure) sobre los diferentes modelos de clasificaci\'on: [Q] basado en un clasificador lineal, [S] y [C] basados en un clasificador SVM. Se usaron 36 eventos (18 crisis and 18 no-crisis) para cada ritmo cerebral; donde $r=1$ es una correlaci\'on total, $r=0$ es una ausencia de correlaci\'on, $r=-1$ es una no-correlaci\'on total. $IC95\%$ es el 95 por ciento del intervalo de confianza. [Q] presenta una alta correlaci\'on para las bandas \emph{delta},\emph{theta}, \emph{alfa} y \emph{beta}, excepto para la banda \emph{gamma} para los eventos de no-crisis, mientras que para los eventos de crisis presenta una ausencia de correlaci\'on. Los modelos [S] y [C] no tienen un correlaci\'on clara. Esto sugiere que nuestro modelo [Q] puede estimar los cambios entre eventos crisis y no-crisis en una escala de $[-1,+1]$.}
\end{table}

La detecci\'on para todos los ritmos cerebrales,  basado en el intervalo de confianza $IC95\%$ del coeficiente de correlaci\'on producto-momento de Pearson de las se\~nales estudiadas es de: 1 segundo en promedio antes del inicio de la crisis para el $50\%$, detecci\'on exacta en tiempo del inicio de la crisis  para el $30\%$ y el $20\%$ restante tiene una latencia en promedio de $3.07$ segundos.

\section{Conclusiones}
\label{sec:disc}
Los resultados preliminares de este trabajo sugieren que la detecci\'on de eventos de crisis utilizando la distribuci\'on Gaussiana generalizada basada en un clasificador lineal junto con 
el coeficiente de correlaci\'on producto-momento de Pearson, permiten predecir una crisis de una no-crisis en un valor de escala entre $[-1,+1]$ en todos los ritmos cerebrales. Por lo tanto, es potencialmente interesante para idear algoritmos de detecci\'on de crisis de manera autom\'atica en tiempo real.

Las perspectivas para el trabajo futuro incluyen una extensa evaluaci\'on de esta metodolog\'ia, mejorar el tiempo de latencia de la detecci\'on de la crisis, realizar un estudio detallado de la confiabilidad de la predicci\'on a medida que se genera la crisis, aplicar nuestra metodolog\'ia en se\~nales a largo-plazo durante el sue\~no, hacer pruebas con se\~nales en tiempo real y comparaciones con otros m\'etodos de predicci\'on del estado del arte, que incluyan una alta robutez frente al ruido y control de artefactos, as\'i como  la intensidad, la duraci\'on y la propagaci\'on de las crisis por cada canal. 

\section*{Conflictos de Interes}
Los autores declaran que no tienen ning\'un conflicto de intereses.

\section*{Agradecimientos}
Parte de este trabajo fue subsidiado por \emph{ITBACyT} DP.557No.34/2015,  Actividades cient\'ificas y tecnol\'ogicas del departamento de investigaci\'on del Instituto Tecnol\'ogico de Buenos Aires y por el protocolo \emph{07/15} del FLENI.


\bibliographystyle{elsart-num}

\end{document}